\documentclass[prl,letterpaper,twocolumn,showpacs]{revtex4-1} 
\usepackage{times,xspace} 
\usepackage{amsbsy,amssymb,amsmath,bm,bbold} 
\usepackage{graphicx,color,epsfig,rotate} 
\usepackage{fancyhdr} 
\usepackage{epstopdf}
\usepackage{float}
\usepackage[colorlinks=true,linkcolor=blue,citecolor=blue,urlcolor=blue]{hyperref}

\def\bbbc{{\mathchoice {\setbox0=\hbox{$\displaystyle\rm C$}\hbox{\hbox 
to0pt{\kern0.4\wd0\vrule height0.9\ht0\hss}\box0}} 
{\setbox0=\hbox{$\textstyle\rm C$}\hbox{\hbox 
to0pt{\kern0.4\wd0\vrule height0.9\ht0\hss}\box0}} 
{\setbox0=\hbox{$\scriptstyle\rm C$}\hbox{\hbox 
to0pt{\kern0.4\wd0\vrule height0.9\ht0\hss}\box0}} 
{\setbox0=\hbox{$\scriptscriptstyle\rm C$}\hbox{\hbox 
to0pt{\kern0.4\wd0\vrule height0.9\ht0\hss}\box0}}}}

\begin{document} 
\title{Towards Hosting Bound State in Continuum in Specialty Optical Microcavity}
\author{Harsh K. Gandhi, Arnab Laha, and Somnath Ghosh}
\email{somiit@rediffmail.com}
\affiliation{Department of Physics, Indian Institute of Technology Jodhpur, Rajasthan-342037, India}

\begin{abstract} 
We exploit the interaction between supported proximity resonances in an open non-uniformly pumped optical microcavity to host a Bound State in Continuum (BIC). Using the modeling of the S-matrix, we study the coupling between the interacting states forming a BIC. We report the divergence of Quality Factor (Q-factor) of one of the interacting states, numerically demonstrated with the appropriate tuning of the spatial variation of unequal gain-loss within the system whose stability has been discussed in terms of the Petermann factor at the point of BIC. Such high lifetime has been marked as a signature of BIC.
\end{abstract} 
 
 
\maketitle %

Bound state in continuum (BIC) can be realized to be waves despite being in the continuum, remain decoupled and hence resultant leakage radiation would be zero. These states are often characterized to be modes having infinite lifetime\cite{stillinger}. Realization of BICs in photonic systems has been an area of extensive research due to its varied application for a new range of devices such as sensors, wave-filters and lasers etc. Lately, there has been substantial analysis towards implementation of BIC in quantum mechanical systems along with its quantitative interpretation for the state to possess infinite lifetime\cite{stillinger}.With this approach to attain infinite lifetime, there have been efforts to obtain such modes by interferences using parameter tuning with the implementation of two parallel dielectric gratings or in an array of waveguides have been reported\cite{marinica}.The principle of parameter tuning was then extended to various coupling resonance scenarios having either single or multiple resonances\cite{hsu}.Due to its varied potential applications in various on-chip devices and other integrated photonic systems, there has been substantial study to host BIC in a variety of basic photonics systems that include photonic crystals, optical waveguides, supperlatice structures with a single impurity potential\cite{capasso} etc. Theoretically the existence of BIC was demonstrated in a pair of quantum dots coupled to reservoirs\cite{guevara} and in 2-D quantum dot pairs connected by quantum lead\cite{ordonez}.With the advent of various models of surface states came the implementation of BIC of a finite periodic system\cite{sprung}, however, to the best of our knowledge, no reports demonstrating the BIC in a versatile platform like optical microcavity are available. Confining light and tuning the coupling inside a cavity has been studied to a very ordinal extent in the recent times. With the application of this pre-requisite knowledge of optical microcavities to confine light lays down the platform to design and execute an ideal fabricated specialty optical microcavity that would confine light infinitely.This light-matter interaction would have its basics with the non-Hermitian quantum formulation of, due to which the amount of energy of each individual mode would not be conserved and hence the eigenstates would be deemed to be complex in nature\cite{moiseyev}. The measure of deviation of this ideality would be captured by the quantity called as the Quality factor (Q-factor)\cite{vahala}. Ideally for an infinite lifetime state enclosed in the optical microcavity would possess infinite Quality. Hence this approach to enhance the state-specific quality of the cavity with the corresponding minimization of excessive noise is the major task to host BIC. In this regards, the concept of avoided resonance crossing\cite{laha,song,wiersig} has shown the potential to physically enhance quality of the interacting states to a very high value.

In this letter, we choose a specialty designed optical microcavity , whose tuning would be optimized using parameter based tuning, in the presence of a weak coupling regime in an unequal gain-loss profile\cite{laha} , we report the presence of a state having ultra-high quality factor , which was identified to be a signature of BIC.The selection of a simple two-port 1D optical microcavity to represent the hosting of BIC and the study of the subsequent features in terms of coupling, stability have been reported which are general and can be extended for any other geometry of cavity. We propose a potential system with higher stability of performance in terms of Q-factor which could aim to host BIC.Further direction to define BIC out of the features obtained haven been established.

Successively, the interaction between the incident states can be explained employing a system of 2X2 Hamiltonian matrix\cite{wiersig}. This matrix would embody the system under consideration which would contain the defined parameters signifying both passive energies of the states i.e. $\varepsilon_1$ and $\varepsilon_2$ along with the corresponding perturbation $V,\,W$ which the system is subjected to.
\begin{equation}
H=\begin{pmatrix}
\varepsilon_1 & 0\\
0 & \varepsilon_2
\end{pmatrix} +
\begin{pmatrix}
0 & V\\
W & 0
\end{pmatrix}=
\begin{pmatrix}
\varepsilon_1 & V\\
W & \varepsilon_2
\end{pmatrix}
\end{equation}
Equation 1. suggests the system under consideration wherein the perturbation is taken off diagonal for the sake of simplicity. Furthermore the eigenvalues of the corresponding states can be obtained which are
\begin{equation}
E_{\pm}=\frac{\varepsilon_1+\varepsilon_2}{2} \pm \sqrt{\frac{(\varepsilon_1-\varepsilon_2)^2}{4}+VW}
\end{equation}
We consider the system to be closed i.e. no external coupling is involved between the states, for which the condition $V=W^*$ sustains. With the existence of this mathematical relation between the perturbating factor, there is a possibility of two types of ARCs \cite{teller}that could physically occurs in the proposed Hamiltonian as described in Eq.1.$|V|^2 > \Im (\varepsilon_1-\varepsilon_2)/2$ would lead to avoided crossing in the imaginary domain along with a crossing in the subsequent real part, which would be driven by the strong coupling that is possible between the passive energies of the system. Conversely $|V|^2 < \Im (\varepsilon_1-\varepsilon_2)/2$, weak coupling would cease to exist, that would lead avoided crossing in the imaginary part and a cross in the real part. In the direction of BIC which are defined to be waves in an open system having zero leakage radiation for which these specialized states are said to have a signature of infinite lifetime or infinity quality factor. Quantum mechanically these states would be described to exhibit zero loss factor i.e. $ \Im[E_\pm]=0$. Further simplification of the underlying condition for an interacting state to exhibit BIC can be expressed as
\begin{equation}
\Im \left(\frac{\varepsilon_1-\varepsilon_2}{2}\right) \pm \left(\frac{\sqrt{\alpha^2 -\beta^2}-\alpha}{2}\right)^{1/2} 
\end{equation}
where,
\begin{eqnarray}\nonumber
\alpha=\frac{[\Re(\varepsilon_1)-\Re(\varepsilon_2)]^2-[\Im(\varepsilon_1)-\Im(\varepsilon_2)]^2}{4}+|V|^2,\\
\nonumber\beta=\frac{[\Re(\varepsilon_1)-\Re(\varepsilon_2)][\Im(\varepsilon_1)-\Im(\varepsilon_2)]}{2}.
\end{eqnarray} 
\begin{figure}[t]
	\centering
	\includegraphics[width=6.5cm]{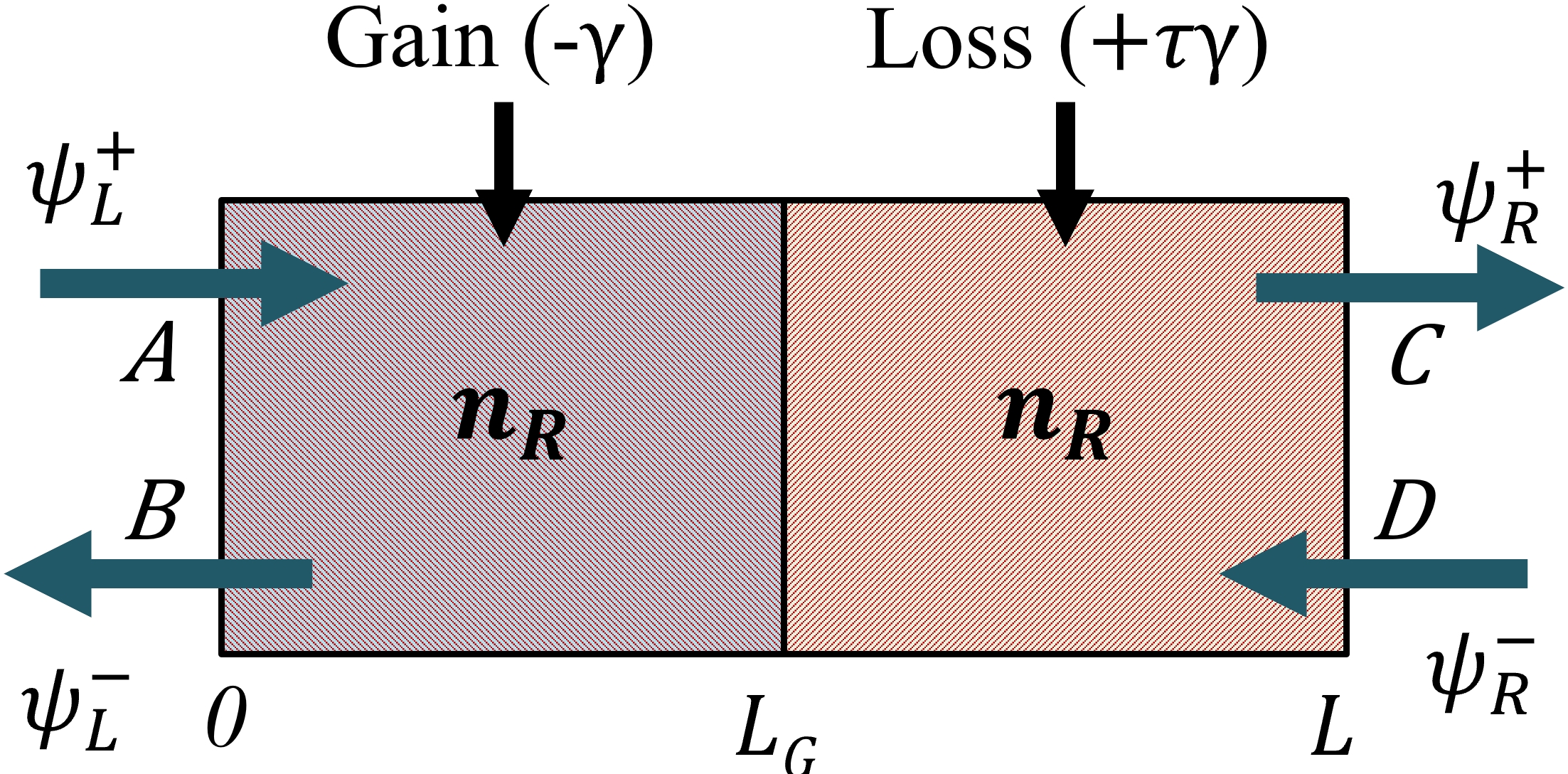}
	\caption{Schematic of two-port network that is specially described in the region $0\le$x$\le$L wherein the optical cavity is divided into two parts having unequal gain($\gamma$) and loss ($\tau \gamma$).Representing $\psi_l^+$ and $\psi_R^-$ as the complex wave-functions for the incident modes(A and D) and $\psi_l^-$ and $\psi_R^+$ as the complex wavefunctions for the scattered modes (B and C) for which we define the scattering matrix taken upon consideration.}
\end{figure}
\begin{figure}[t]
	\centering
	\includegraphics[width=6.5cm]{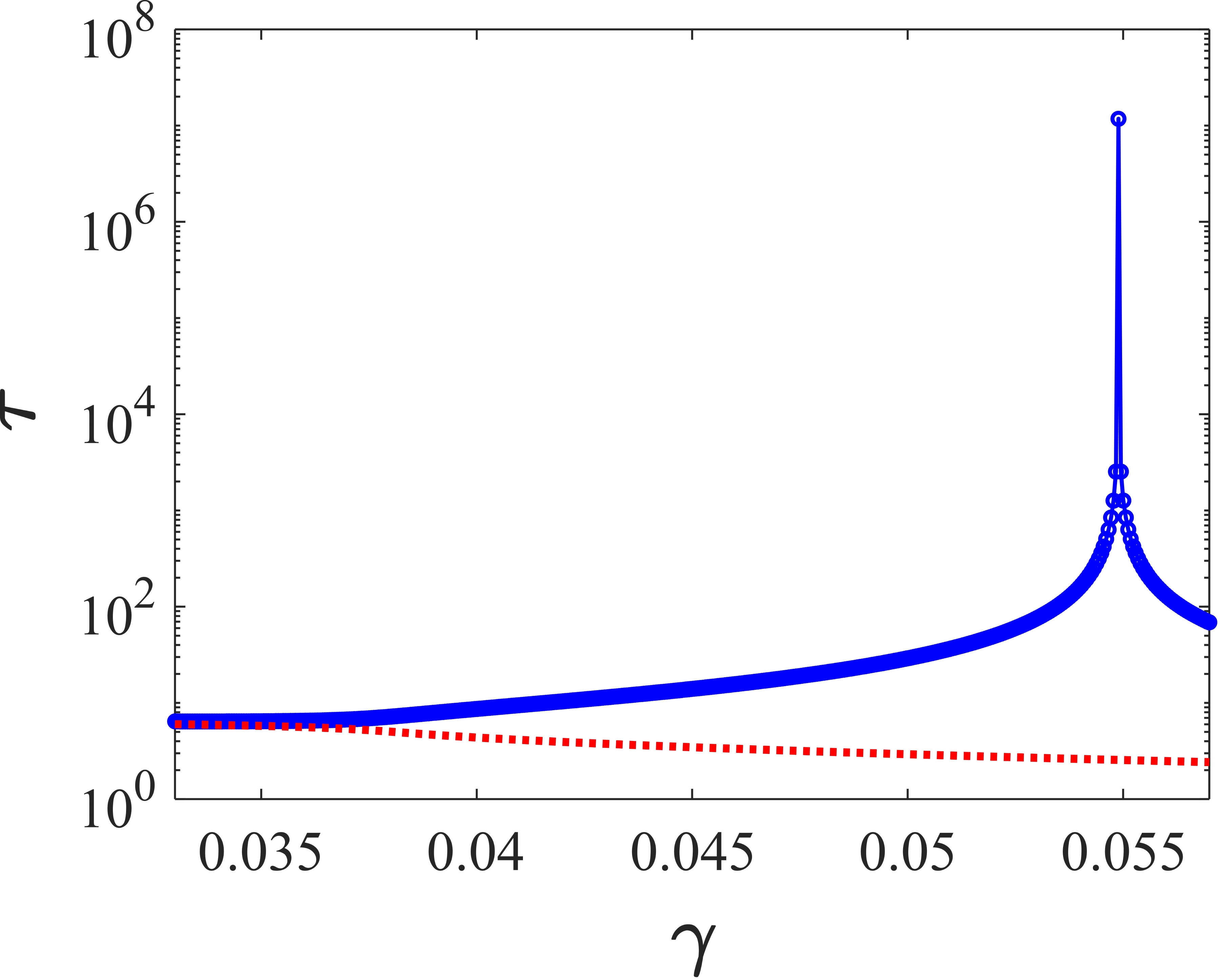}
	\caption{Variation of the lifetime of the states formed due to the mutual coupling of the incident states with the amount of gain injected in the designed system. The blue solid line represents the variation of the lifetime of the longer living state which exhibits maximum lifetime of $1.17$x$10^7$ a.u. at a gain of 0.05489 which is later identified to the signature of the existing BIC Subsequently the red-dotted line represents the lifetime of the shorter living decaying state having a lifetime of 2.556 a.u. at the same amount of gain. This enhancement of one prescibed state along with the decay of the cease to exist after the ARC gain of 0.037.}
\end{figure}
\begin{figure}[t]
	\centering
	\includegraphics[width=6.5cm]{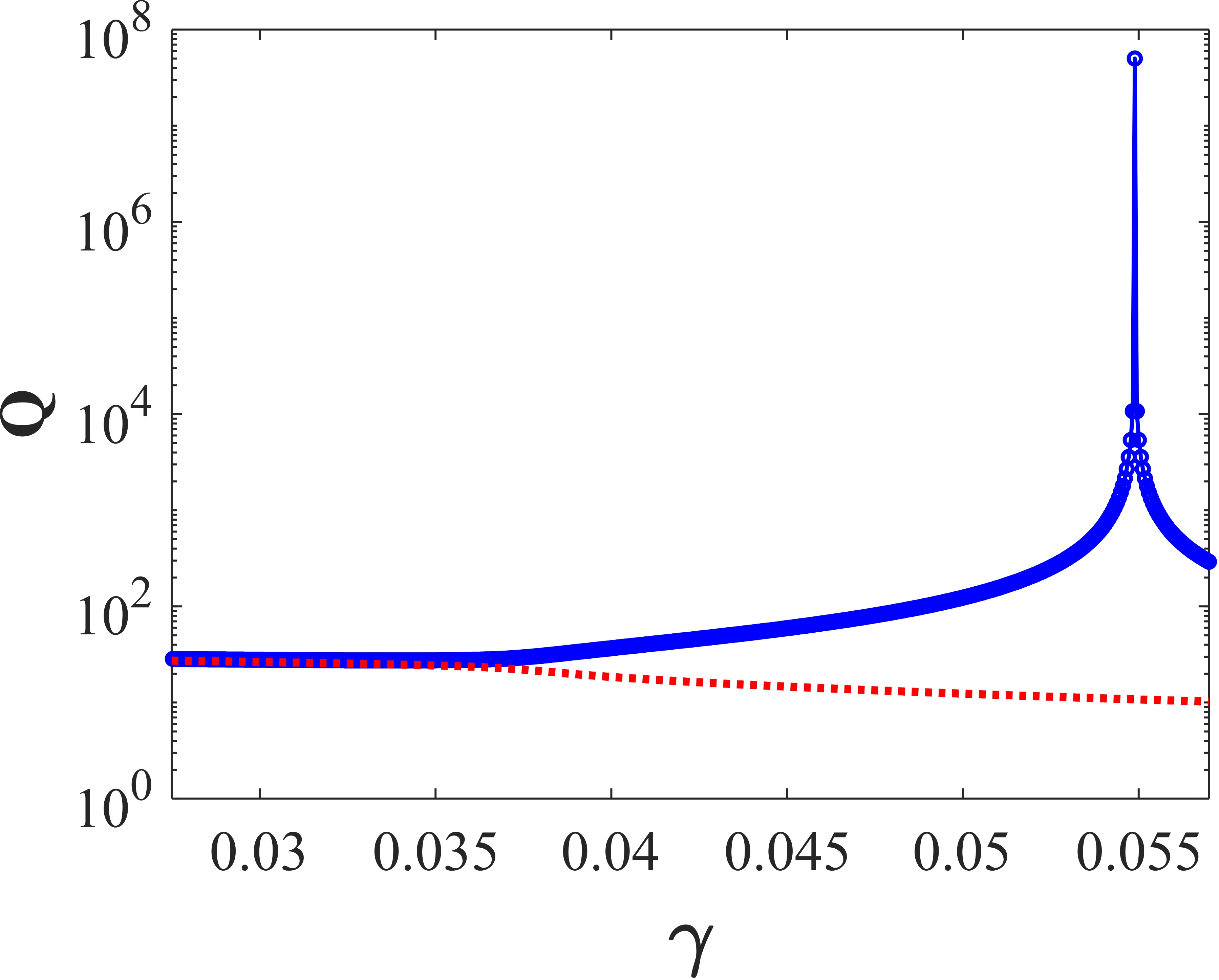}
	\caption{Variation of the Q-factor(Q) of the states formed with the amount of gain injected in the designed system. This variation would follow the same nature of plot of lifetime.}
\end{figure}

The numerical modeling of the same was taken into consideration to look into various profiles that could provide the evidence of hosting of BIC in a microcavity with appropriate tuning and perturbation. We consider a 1D two port optical microcavity with Fabry-Perot geometry. We could also considered toroidal resonator or any other resonator with proper phase matching but for the sake of simplicity to implement the coupling along the axis of cavity, hence justifying the choice of selection of cavity under consideration. The proposed cavity of length $L$ and a background refractive index of $n_R$ is divided into two parts, one occupying the region $0\leq x \leq L_G$ designated as the gain region within which a measure of gain of $\gamma$(coefficient of gain) is injected, and the other region extending from $L_G \leq x \leq L$ designated as the loss region having a loss of $\gamma\tau$ incorporated into it.
Hence the primary elementary refractive index of both the parts $n_G,n_L$ can further be written with the inclusion of gain/loss as $n_G=n_R-i\gamma$ and$n_L=n_R+i\tau\gamma$. For our proposed system we take the length of both the parts to be equal to 5$\mu$m and an optimized value of $\tau$ to be 1.59 with a motivation to make our system overall loss dominating.We choose the value of $n_R$ to be 1.5 for sake of practicality and for ease of fabrication.

Further the scattering matrix (S-matrix) of our proposed speciality optical microcavity can be described as 
\begin{equation}
\left[\begin{array}{c}B\\C\end {array}\right]=S(n(x),\omega)\left[\begin{array}{c}A\\D\end {array}\right];
\label{equation_s-matrix}
\end{equation}
\begin{figure}[t]
	\centering
	\includegraphics[width=6.5cm]{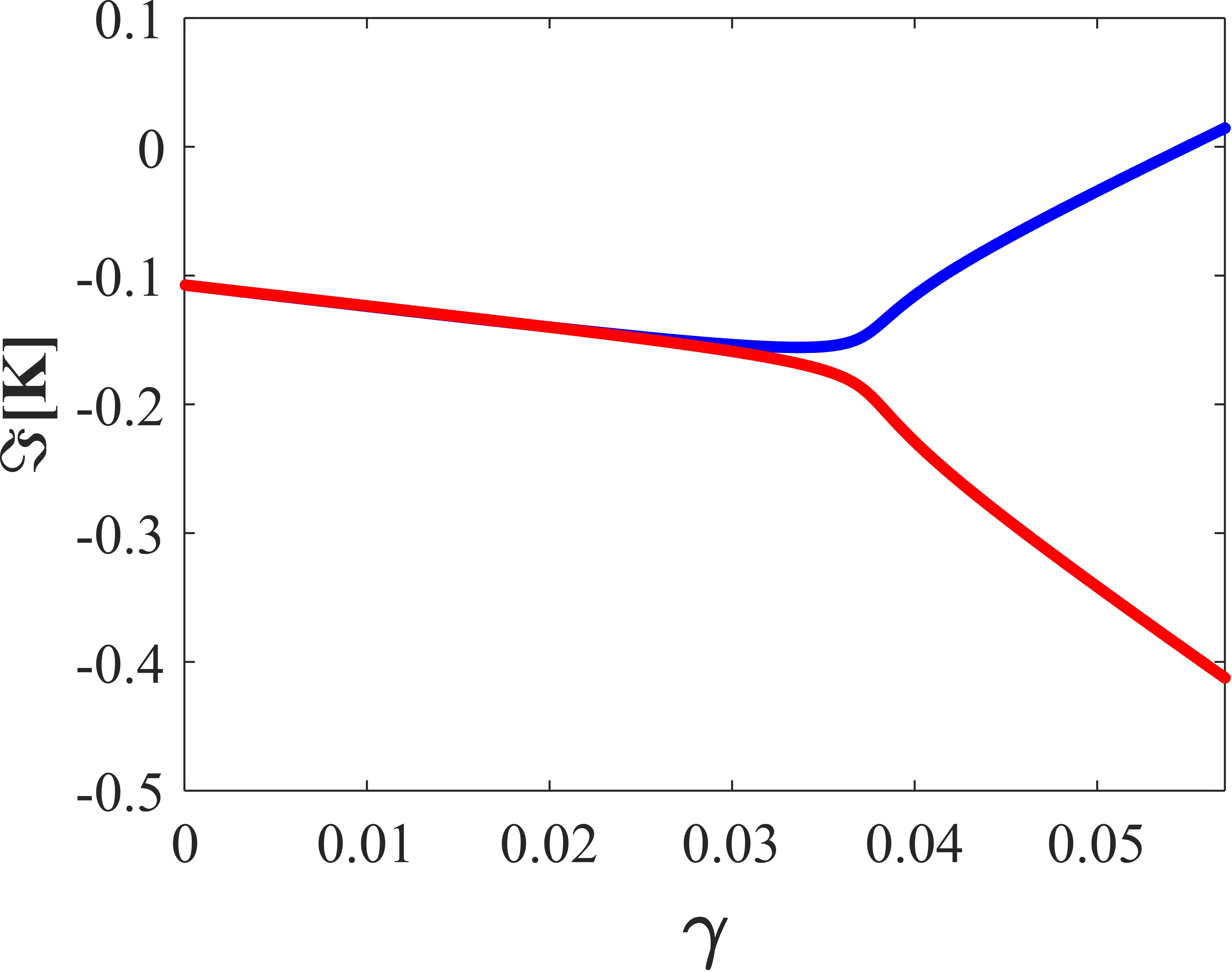}
	\caption{Variation of the imaginary part of the wavenumber of the coupling modes wherein the red curve depicts the mode with shorter lifetime and the blue curve depicts the mode with the longer lifetime.}
\end{figure}
\begin{figure}[ht]
	\centering
	\includegraphics[width=6.5cm]{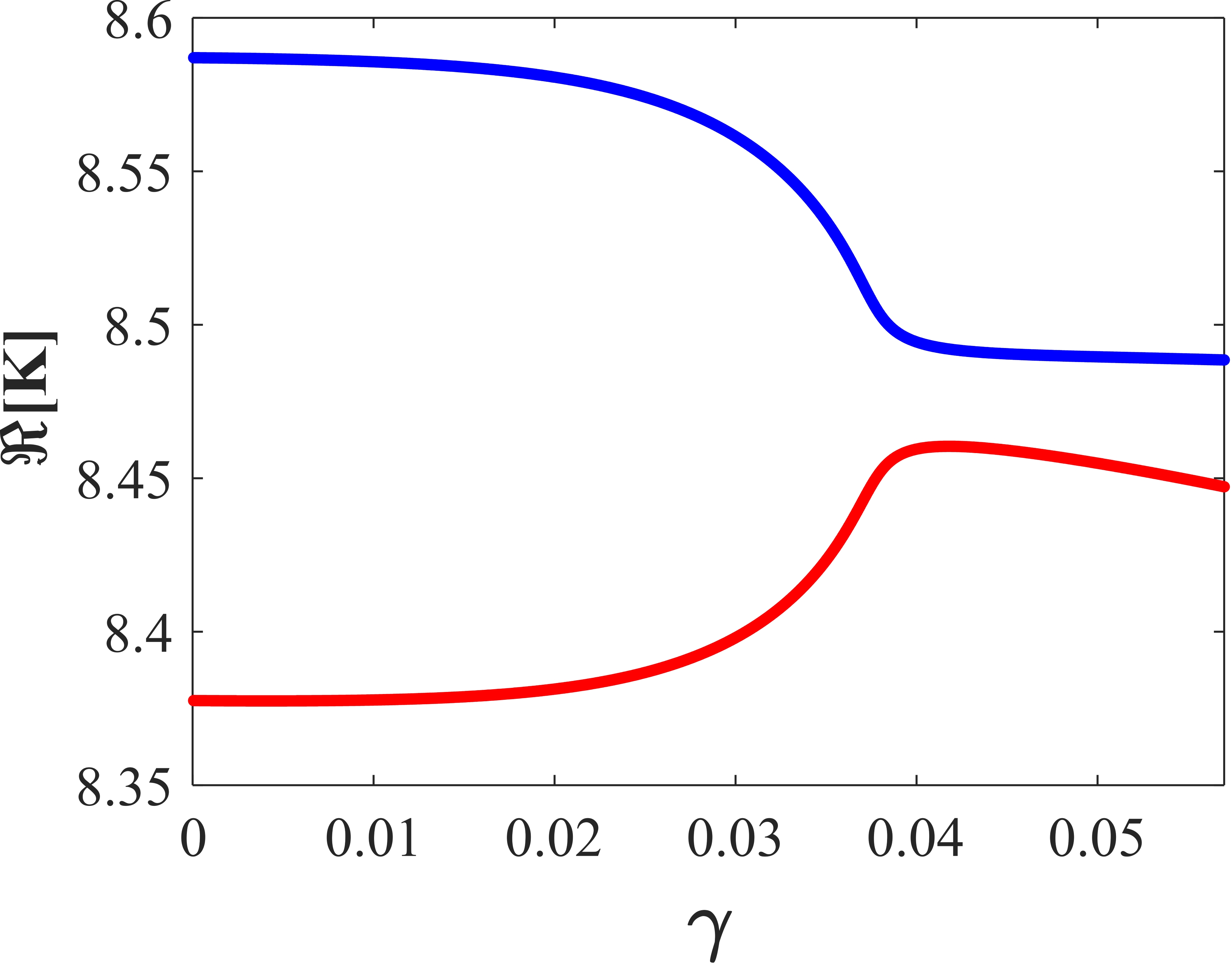}
	\caption{Variation of the Real part of the wavenumber of the coupling modes wherein the red curve depicts the mode with shorter lifetime and the blue curve depicts the mode with the longer lifetime.Plot shows no visible crossing of the real part of the wave numbers of the interacting states.}
\end{figure}
where A,D are the amplitude of the incident waves and B,C are the amplitude of the scattered wave output from the left and right side of the cavity.Physical realization of the solutions of Eq.3 can only be considered with the utilization of both complex and real parts of the frequency component.For which we find the poles of the S-matrix using the root finding numerical tool on 
\begin{equation}
\frac{1}{\max[eig(S(\omega))]}=0
\end{equation}

The modes contemplate the transient trapping of the scattering particle and hence would be equivalent to the poles of the scattering matrix as found in Eq.5. To understand the interaction leading to BIC, primarily we locate two arbitrary poles of S-matrix for any normalized standard frequency which lie in between 8.3 and 8.7 (in $\mu m^{-1}$) and the imaginary part lies in between -0.05 to 0.05. We then show the formation of a long and short lifetime states resultant after coupling and the mutual exchange of energy by plotting both the lifetime and quality factor with respect to increasing gain at a fixed $\tau$. Mathematically we profuse lifetime to be $1/2\Im(k)$ and quality factor(Q-factor) to be $\Re(k)/2\Im(k)$ where $k$ is wave number of the interacting modes. 

We report that the lifetime of one of the modes interacting that scales up that is essentially identified to have the features to be designated to be BIC. Theoretically for a BIC that would exhibit no leakage loss, to imply that the $\Im[k]$ would tend to zero or the lifetime/ Q-factor would tend to infinite. Figure 3 and 4 clearly represent the above said signature of BIC. While the shorter lived state (red -dotted line representing lifetime in figure 2 and Q-factor in figure 3) experiences an very drastic decay in both the qualities in consideration, the longer lived state(blue  line representing lifetime in figure 2 and Q-factor in figure 3) claiming to be BIC, would reach its peak at $\gamma=0.05489$ for the prescribed preset of $\tau$. We further deduce that at the point of BIC i.e. gain coefficient is 0.05489, the longer living state survives for $\approx 1.17$x$10^7$ having a quality of $\approx 5$x$10^7 $ , with a subsequent lifetime and Quality of 2.556 and 10.8 of the decaying mode coupled with the mode exhibiting BIC.An important point to observe is the formation of such high quality states occurs in a cavity that is overall lossy.With the establishment of the state having very high Q-factor as BIC, we further study the nature of interaction between the states and at that particular value of $\tau$ understand the extent of coupling between the interacting states.

The beginning of the understanding would be the identification of the location of the ARC, which is in the parameter space of $(\gamma,\tau)$ designated as $\gamma_c$ and $\tau_c$ at which the real or imaginary parts of the wave numbers of the interacting modes would be equal.Figure 4 and 5 show the plot of the real and imaginary part of the wave numbers where the blue and red lines represent the longer and shorter living modes respectively. With the preset of the already set value of $\tau$ we show the anti crossing in the real part of the wave numbers and a subsequent crossing in the imaginary part of the same.We illustrate that at the value of $\gamma \approx 0.037$ and $\Im[k] \approx -0.2$ there appears to be a crossing in the imaginary part and at $\Re[k] \approx 8.47$ there seems to be no visible crossing in the real part hence suggesting weak coupling between the states. Essentially the decay of the shorter living state in the Q-factor and lifetime occurs near the ARC gain coefficient or $\gamma_c$. 

It can be observed that the state of BIC appears does not cease to exist at the point of ARC. Furthermore, to explore the features of BIC the variation of the eigenfunction of the state designated as BIC(the longer living state) was plot in Figure 6. The plot illustrated a significant dip which could also be thought quantum mechanically as the continuum edge of the potential well mode of BIC. For a bound state to exhibit the characteristics in the continuum region would have to cross the continuum edge. At the above found point of BIC i.e. $ \gamma= 0.05489$ the value of amplitude of wavefunction would tend to 1. Hence this state existing would be said to be a closed/ bound system existing in the parameter region beyond the continuum state which would further be a bound state in the continuum region.
\begin{figure}[ht]
	\centering
	\includegraphics[width=6.5cm]{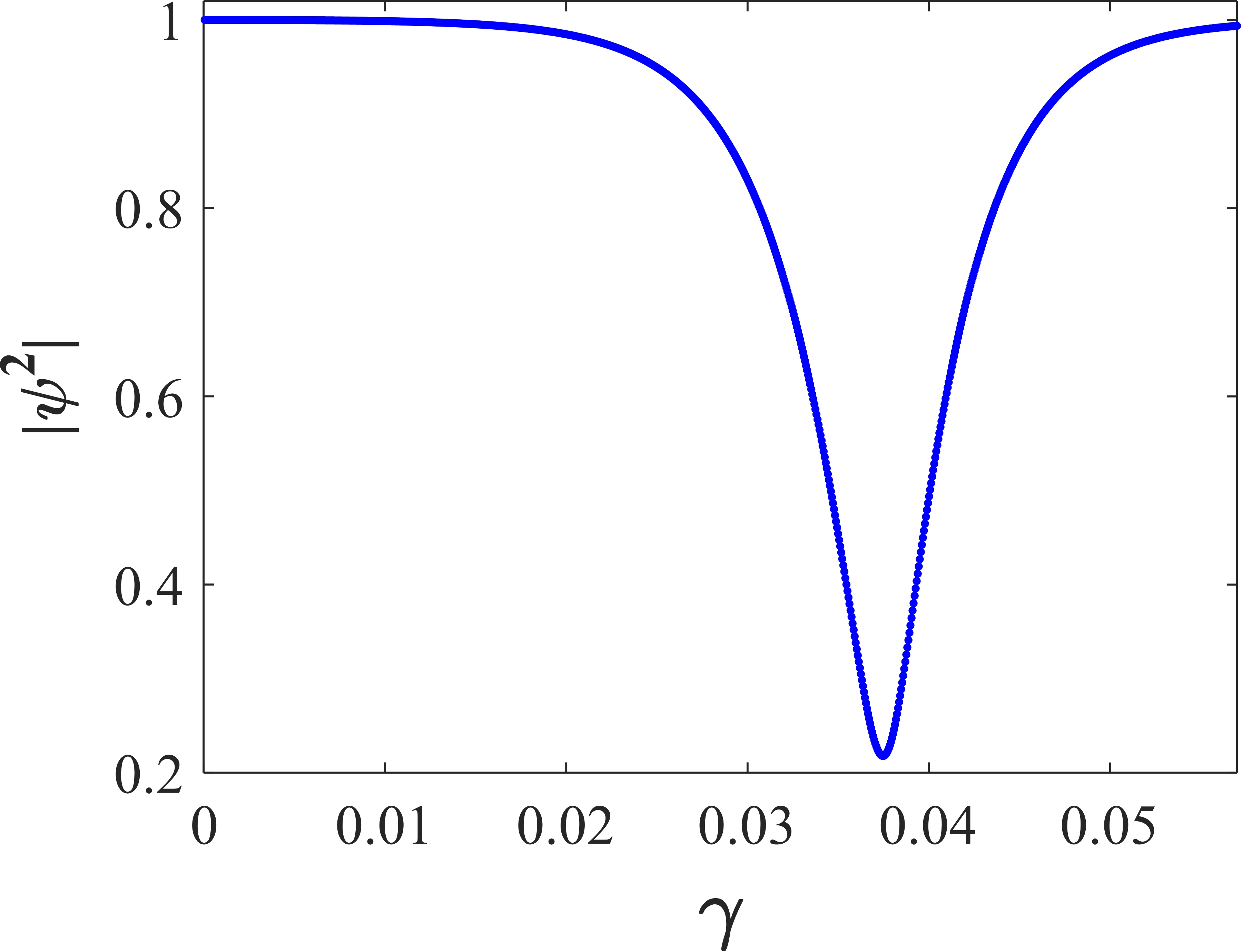}
	\caption{Variation of amplitude of complex wavefuntion with the increasing gain. }
\end{figure}
\begin{figure}[ht]
	\centering
	\includegraphics[width=6.5cm]{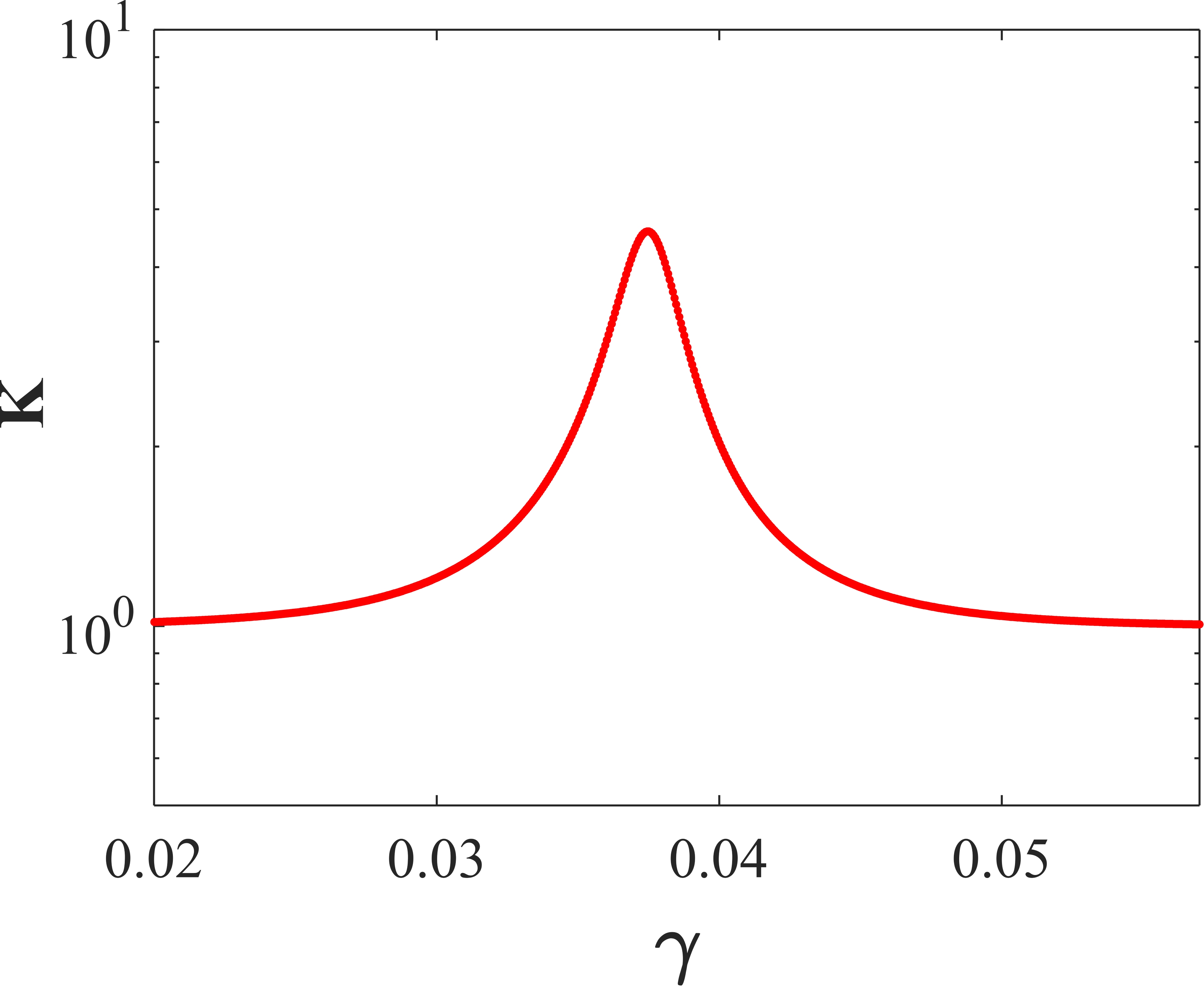}
	\caption{Corresponding variation of the Petermann factor with gain of the longer living mode.}
\end{figure}

If the wavefunction of the $n^{th}$ mode is $\psi_n$ then the {\it Petermann} factor\cite{petermann} $K_n$ is expressed as
\begin{equation}
K_n=1/|\langle \psi_n^*|\psi_n\rangle|^2
\end{equation}
This factor is an excellent indicator of the noise incurred due to the non-orthogonality of the coupled states along any proposed geometry of resonator. With the plot of variation of the gain with K of the longer living mode, we find that the value of K at the point of BIC is said to be very close to 1. This K would be having its peak at the point of dip of the wavefunction of the corresponding state as the interacting modes traverse faster when they approach the ARC point.Hence at the point of BIC it can be established that a stable state having theoretically infinite lifetime, away from the point of crossing would cease to exist.

In summary, we have described, with an adequate amount of tuning of the injected unequal gain and loss profile of an optical microcavity could lead to the enhanced quality factor for a longer living state with a subsequent decay of the quality factor of the corresponding counterpart.Moreover, this state having very high lifetime which is to realize BIC in such photonic systems.The features of BIC have been discussed along with their implementation and its stabilization in terms of performance with a reduced Petermann factor at the point of BIC.Such specially designed microcavities whose stability due to the presence of state-of-art micro fabrication and implementation techniques could play a vital role of a suitable host of BIC whose stability would be less susceptible to the fabrication techniques.

\vspace{0.5cm}
A.L. and S.G acknowledge the financial support from the Science and Engineering research Board (SERB) [Grant No. ECR/2017/000491], Department of Science and Technology, Government of India.

\end{document}